\documentclass[12pt,preprint]{aastex}
\usepackage[twocolumn,numberedappendix]{emulateapj5}

\usepackage{epsfig}
\usepackage{rotating}
\tightenlines
\slugcomment{Accepted to the Astrophysical Journal Letters}

\newcommand \lsim{\mathrel{\rlap{\lower4pt\hbox{\hskip1pt$\sim$}}
    \raise1pt\hbox{$<$}}}
\newcommand \gsim{\mathrel{\rlap{\lower4pt\hbox{\hskip1pt$\sim$}}
    \raise1pt\hbox{$>$}}}

\newcommand     \kms    {\,{\rm km~s}^{-1}}

\newcommand{\smyr}{{ M_\odot\ \rm yr^{-1}}}
\newcommand{\sm}{{M_\odot}}

\newcommand{\beq}{\begin{equation}}
\newcommand{\eeq}{\end{equation}}
\newcommand{\beqa}{\begin{eqnarray}}
\newcommand{\eeqa}{\end{eqnarray}}

\newlength{\figwidth}
\countdef\decade=200
\advance\decade by \year
\countdef\hours=201
\advance\hours by \time
\divide\hours by 60
\countdef\mins=202
\advance\mins by \hours
\multiply\mins by 60
\multiply\hours by 100
\countdef\miltime=203
\advance\miltime by \hours
\advance\miltime by \time
\advance\miltime by -\mins

\begin{document}

\title{The Becklin-Neugebauer Object as a Runaway B Star,\\Ejected 4000 years ago from the $\theta^1C$ system}

%\centerline{DRAFT: \today}

\author{Jonathan C. Tan}
%\footnote{Princeton University Observatory, Princeton, NJ 08544, USA, jt@astro.princeton.edu}}
        
\affil{Princeton University Observatory, Peyton Hall, Princeton, NJ
08544, USA.\\jt@astro.princeton.edu}

\begin{abstract}
  We attempt to explain the properties of the Becklin-Neugebauer (BN)
  object as a runaway B star, as originally proposed by Plambeck et
  al. (1995). This is one of the best-studied bright infrared sources,
  located in the Orion Nebula Cluster --- an important testing
  ground for massive star formation theories.  From radio observations
  of BN's proper motion, we trace its trajectory back to Trapezium
  star $\theta^1C$, the most massive ($45\sm$) in the cluster and a
  relatively tight (17~AU) visual binary with a B star secondary.
  This origin would be the most recent known runaway B star ejection
  event, occurring only $\sim4000\:{\rm yr}$ ago and providing a unique
  test of models of ejection from multiple systems of massive stars.
  Although highly obscured, we can constrain BN's mass ($\simeq
  7\sm$) from both its bolometric luminosity and the recoil of
  $\theta^1C$. Interaction of a runaway B star with dense ambient
  gas should produce a compact wind bow shock. We suggest that
  X-ray emission from this shocked gas may have been seen by {\it
    Chandra}: the offset from the radio position is $\simeq300$~AU in the
  direction of BN's motion.  Given this model,
  we constrain the ambient density, wind mass-loss rate and
  wind velocity. 
%The density is further constrained by
%  observations of the density of shocked neutral gas near BN, while
%  the X-ray spectrum suggests a relatively fast wind velocity. 
%  We describe the necessary conditions to achieve a self-consistent
  %model. 
  BN made closest approach to the massive protostar, source ``I'',
  500~yr ago. This may have triggered enhanced accretion and thus
  outflow, consistent with previous interpretations of the outflow
  being a recent ($\sim 10^3$~yr) ``explosive'' event.
\end{abstract}


\section{Introduction}\label{S:intro}

The Becklin-Neugebauer (BN) object was discovered as a bright $2\:{\rm
  \mu m}$ source (Becklin \& Neugebauer 1967), about 0.1~pc
(45\arcsec) in projection from the Trapezium stars of the Orion Nebula
Cluster (ONC) (e.g. O'Dell 2001), about 450~pc distant
(Genzel \& Stutzki 1989).  It was initially thought to be a protostar
(i.e. embedded and in main accretion phase), since it is
infrared bright but optically faint. The extinction-corrected
spectrum peaks at $\sim5\rm \mu m$ ($T\simeq 550\:{\rm K}$), with
$A_V\simeq 17$~mag (from depth of silicate $9.8\:{\rm \mu m}$
absorption), yielding a bolometric luminosity $\sim 2500L_\odot$ from
a ``core'' region ($\sim 1\arcsec$) and $\sim 10^4L_\odot$ from
an extended region ($\sim 10\arcsec$) (Gezari, Backman, \&
Werner 1998). These estimates are uncertain as they are derived from
single temperature blackbody fits to limited portions of the spectrum:
as Gezari et al. describe, the BN region probably exhibits a range of
temperatures.  The extinction correction is also uncertain: the above
estimate agrees with Scoville et al.'s (1983) determination from the
Br$\alpha$/Br$\gamma$ emission line ratio assuming case B
recombination theory.  However Bunn, Hoare, \& Drew (1995) argued
from the Pf$\gamma$ line strength that case B theory is not
applicable and $A_V>18$~mag by a significant amount.
%This would raise the estimate of BN's luminosity.

If the $2500-10^4 L_\odot$ corresponds to the luminosity of a solar
metallicity zero age main sequence star, then this has mass $m_*\simeq
8.4 - 12\sm$, radius $r_*\simeq 3.4-4.3\:R_\odot$ and surface
temperature $T_*\simeq 22,000-28,000\:{\rm K}$ (Schaller et al. 1992).
BN has 8.3~day periodic variability with $\sim 0.2$~mag peak to trough
change at H and K (Hillenbrand, Carpenter \& Skrutskie 2001). This
behavior is similar to that of JW~660, a mid-B spectral type ($\sim
6\sm$) star also in the ONC with a 6.15~day period (Mandel \& Herbst
1991). These periods are somewhat longer than those typical of slowly
pulsating B stars (Mathias et al. 2001).

BN is detected in the radio (Moran et al. 1983; Garay, Moran \& Reid
1987; Churchwell et al. 1987; Menten \& Reid 1995; Plambeck et al.
1995) and just resolved at 2~cm with diameter $31 \times 18$~AU
($0.07\arcsec \times 0.04\arcsec$) (Churchwell et al. 1987).  The
spectrum rises as $\nu^{1.25}$ to $\sim 100$~GHz and then may
flatten. Plambeck et al. (1995) present a non-unique model of this
emission: a $10^4$~K \ion{H}{2} region with uniform density core
($n_e= 5\times10^7\:{\rm cm^{-3}}$; radius $r_0=7$~AU) and steeply
declining envelope with $n_e\propto (r/r_0)^{-3.3}$.  
%However other
%models are also possible.
%, either due to a stellar wind or ionization of ambient gas by the star.

\section{The Motion of BN}\label{S:motion}

From radio data with a time baseline of 8.7 years (1986.3 - 1995.0),
Plambeck et al. (1995) reported a proper motion of BN relative to
source ``I'' (a massive protostar forming from the Orion
hot core, Greenhill et al. 1998; Tan 2003) of
$\sim0.02\arcsec\:{\rm yr^{-1}}$ ($\sim 50\kms$ in the sky plane
for BN at 450~pc, adopted throughout), towards the NW. 
From the four data points of Plambeck et. al., we
find $0.021\pm 0.004\arcsec\:{\rm yr^{-1}}$ ($45\pm8\kms$) towards
position angle (P.A.) $-33\pm 11^\circ$, where errors are $1\sigma$ assuming
the individual coordinates are known to $\sim0.02\arcsec$.  The
relative positions have continued to be determined
with BIMA (Berkeley-Illinois-Maryland-Array)
(Plambeck 2004, private comm.).  From a joint analysis of the
full data set (extending to 2004) we estimate a proper motion of
$0.0181\pm0.0022\arcsec\:{\rm yr^{-1}}$ ($38.7\pm 4.7\kms$)
towards P.A. $-37.7\pm5^\circ$.  Following the velocity
vector backwards assuming no acceleration, we see BN was near the
Trapezium stars $\sim4000$~yr ago (Fig.~\ref{fig:prop3}) and made
a close ($\lesssim 2\arcsec$) projected passage of source ``I''
$\sim500$~yr ago.

Scoville et al. (1983) observed several spectroscopic tracers of BN
and its circumstellar nebula, reporting a velocity relative to the
local standard of rest of $+21\kms$. This is significantly larger than
the ONC mean ($+8\pm3.5\kms$, Walker 1983) and the gas of the Orion
hot core and larger scale molecular cloud ($+9\kms$, Genzel \& Stutzki
1989). Thus relative to the Orion cloud and star cluster, BN is moving
at $v_{\rm BN} \simeq 41 \pm 6\kms$.

If BN originated near the Trapezium center then it
is now $\sim 0.05$~pc further away from us. BN's extinction
($A_V\gtrsim 17$, Gezari et al. 1998, Bunn et al. 1995), implies a
large intervening column of gas and dust, $N_{\rm H}\gtrsim 6\times
10^{22}\:{\rm cm^{-2}}$, 90\% of which is behind the Trapezium stars.
Wen \& O'Dell (1995) considered the creation of an \ion{H}{2} region
by $\theta^1C$ and its interaction with the nearby neutral gas.  They
argue this gas is $\sim0.15$~pc behind the Trapezium stars for a
uniform density model and somewhat closer for exponential
models.  The possible discrepancy of these estimates with the
$\sim$0.05~pc inferred from BN's kinematics warrants further study.
For a neutral layer thickness of $\lesssim 0.05$~pc, the mean density
on the BN sight line is $\gtrsim 4\times 10^{5}\:{\rm cm^{-3}}$.
This material is likely associated with the Orion hot core that
harbors source ``I''.  This core may have had an initial mass $\sim
60\sm$ (Tan 2003) and density at its surface $n_{\rm H,s}\simeq
10^6\:{\rm cm^{-3}}$ (McKee \& Tan 2003), in reasonable agreement with
the above estimate.% from BN's extinction and the geometry of its position.
%The density at the surface of such a core that is in
%equilibrium with a bounding pressure typical of massive star forming
%regions with surface densities of $\Sigma=1\:{\rm g\:cm^{-2}}$
%(equivalent to $P\simeq 0.85\times 10^9\:{\rm K\:cm^{-3}}$; McKee \&
%Tan 2003) is $n_H\simeq 1.1\times 10^6(M_c/60\sm)^{-1/2}\Sigma^{3/2}
%\:{\rm cm^{-3}}$

\section{Creation of a Runaway: Ejection from $\theta^1$C}\label{S:progenitor}

Two models are thought to explain runaway OB stars.
Zwicky (1957) and Blaauw (1961) proposed that they originate when
a supernova occurs in a close binary.
%The
%supernova theory has several problems to explain all runaways: most
%massive close binaries are not expected to disrupt after the explosion
%because of mass transfer from the primary to the secondary (Kornilov
%\& Lipunov 1984), contrary to the observation that most runaways are
%single stars (Gies \& Bolton 1986); the delay in ejection of the
%secondary, after the primary has undergone its stellar evolution,
%reduces the time available for the runaway to reach the observed large
%distances from the plane; and the model is not able to explain the
%observed inverse correlation of velocity of the runaway with its mass
%(Gies \& Bolton 1986).
Alternatively, Poveda, Ruiz, \& Allen (1967) hypothesized 
origin by dynamical ejection of a star from a multiple
system. 
%Hut \& Bahcall (1983) and Hut (1984) considered the general
%problem of binary-single star scatterings, where all stars had equal
%masses.  
%Leonard \& Duncan (1988; 1990) simulated star clusters including
%the effects of binary-binary interactions, and specifically studied
%the production of runaway stars.
%showed that typical open clusters can eject runaways as a
%result of binary-binary interactions, if the binaries are initially
%relatively hard. Velocities up to $200\kms$ were achieved, and
%typically the lowest mass stars had the highest velocities.
Hoogerwerf, de Bruijne \& de Zeeuw (2001) considered past
trajectories of a large sample ($\sim 50$) of nearby ($\lesssim
1$~kpc) runaways and several pulsars. They found one case of a
pulsar and B star that supports the supernova scenario. They
also identified two B stars and a massive eccentric binary
with an ejection event from the ONC $\sim2.5$~Myr ago, which
supports dynamical ejection (see also Gualandris, Portegies-Zwart
\& Eggleton 2004).  \footnote{This event implies that massive star
  formation has been occurring in the ONC for $\sim2.5$~Myr, and
  that a large fraction (almost 50\%) of massive ($m_*>10\sm$) stars
  have been ejected. As Hoogerwerf et al. (2001) point out, this
  affects determinations of the high-mass initial mass function from
  studies of star clusters.
%(e.g. Muench et al. 2002). 
  It also has implications for models of star cluster formation (e.g.
  Tan \& McKee 2001), which assess the importance of feedback
  from massive stars: this is reduced if such stars are
  efficiently ejected from clusters.}

BN, if created in the ONC, must represent an example of dynamical
ejection, since the cluster is young ($\lesssim 3\:{\rm Myr}$; Palla
\& Stahler 1999) and there is no evidence for a recent supernova.
Dynamical modeling (e.g. Leonard \& Duncan 1990) suggests the runaway
should often leave behind a hard, eccentric binary of much greater
total mass and an escape velocity from a location typical of the
secondary's orbit that is comparable to the ejection velocity.

\begin{figure}[h]
\begin{center}
\epsfig{
%prop3
        file=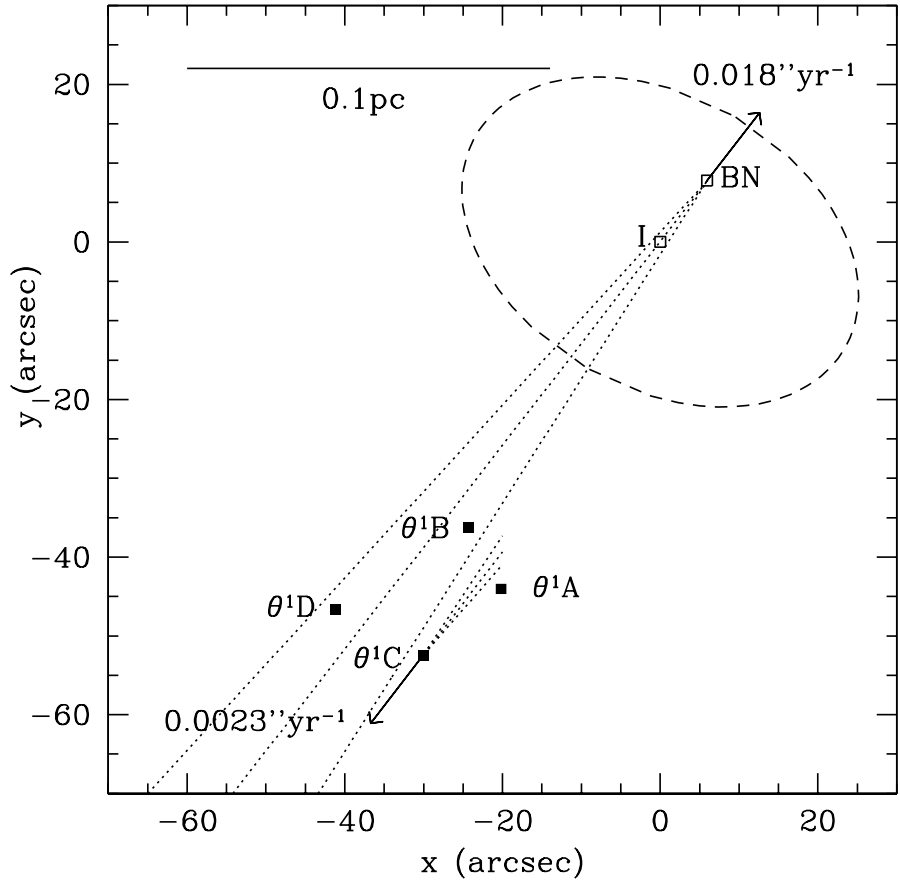,
        angle=0,
        width=\figwidth
}
\end{center}
\caption{ \label{fig:prop3} The velocity vector and inferred past 
  trajectory of BN (central dotted line), based on radio observations
  (Churchwell et al. 1987; Menten \& Reid 1995; Plambeck et al. 1995;
  Plambeck 2004, private comm.).  Coordinates are relative to source
  ``I''. An approximate estimate of the $1\sigma$ uncertainty in the
  position angle of the trajectory is shown by the outer dotted lines.
  BN was near the Trapezium stars, about 4000~yr ago. In particular,
  we argue (see text) that it was ejected from the $\theta^1C$ binary
  system, the proper motion of which is also shown. Note that BN made
  closest passage $500$~yr ago to source ``I'', a massive protostar.
  The dashed line shows the approximate size of an initial equilibrium
  $60M_\odot$ core (McKee \& Tan 2003; Tan 2003), bounded by a
  (nonthermal) pressure $\sim G\Sigma^2$ with $\Sigma=1\:{\rm
    g\:cm^{-2}}$.}
\end{figure}

%Preibisch et al. (1999) looked for binary companions to 13 O and B
%stars in the ONC, including the Trapezium stars. 
All the Trapezium stars lie close to the possible past trajectories of
BN (Fig. \ref{fig:prop3}). Their binary properties are (Preibisch et
al. 1999; Schertl et al. 2003): $\theta^1A$ (B0, $16\sm$) has a visual
companion at 100~AU ($4\sm$) and a spectroscopic companion at $\sim
1$~AU ($\sim 3\sm$); $\theta^1B$ ($7\sm$) has a spectroscopic
companion at $<0.1$~AU ($\sim 2\sm$) and 3 other more distant
companions of at most a few solar masses; $\theta^1C$ (O6, $45\sm$)
has a visual secondary at 17~AU ($\gtrsim 6\sm$), perhaps on an
eccentric orbit (Schertl et al. 2003) and able to explain evidence for
a spectroscopic companion (Donati et al.  2002; Vitrichenko 2002);
$\theta^1D$ (B0.5) has no detected companions.  We note $\theta^2A$
(09.5, $25\sm$), the second most massive ONC star and located at
(-130\arcsec, -150\arcsec) relative to source ``I'' , also lies along
an extrapolation of BN's past trajectory.  It has a spectroscopic
companion at $0.47$~AU ($\sim 9\sm$, $e=0.33$) and a visual companion
at 173~AU ($7\sm$) (Preibisch et al 1999; Abt, Wang, \& Cardona 1991).
The escape velocity from the $\theta^1C$ system from a location
characteristic of the secondary's orbit is $67
(m_{\theta^1C}/50\sm)^{1/2}(r/20{\rm AU})^{-1/2}\kms$, while it is
$\sim 350\kms$ from the inner $\theta^2A$ system. These
appear to be the only ONC stars capable of ejecting BN.

If no other star was ejected with BN, then the parent system should be
recoiling with a proper motion, evaluated for the case of $\theta^1C$,
of $3.6\:{\rm mas\:yr^{-1}} (m_{\rm BN}/10\sm)
(m_{\theta^1C}/50\sm)^{-1}$, where $1\:{\rm mas}\equiv0.001\arcsec$.
This is several times larger than the dispersion of
bright ONC stars ($\pm 0.70\pm0.06 \:{\rm mas/yr}$, van Altena et al.
1988), and much larger than the velocity expected if $\theta^1C$ were
in equipartition with the other cluster stars.
%Of course if BN
%is only one of two stars ejected in the same event, then the motion of
%$\theta^1C$ could be practically zero, although then there would be an
%additional fast star moving to the SE.  
In fact van Altena et al. (1988) report a proper motion of $\theta^1C$
of $2.3\pm0.2$~mas/yr towards P.A. $+142.4\pm4^\circ$ (corrected to
J2000), with the quoted errors resulting from assuming $\pm
0.15$~mas/yr in the x and y velocity vectors (van Altena et al. 1988).
The uncertainties in $\theta^1C$'s motion are $\sim20$\% larger due to
its brightness (van Altena 2004, private comm.), but this does not
significantly change our results. $\theta^2A$ is moving towards
$-30^\circ$ at $\sim2.7$~mas/yr, i.e. roughly towards the Trapezium
and BN, so it is unlikely to have ejected BN.  Note that these motions
are relative to a frame defined by the mean proper motion of van
Altena et al.'s sample of ONC stars, while BN's motion has been
measured relative to source ``I''.  We are assuming that ``I'' and the
ONC have negligible relative proper motion. The direction of
$\theta^1C$'s motion is consistent with being exactly opposite to BN's
(Fig. \ref{fig:prop3}).  We searched van Altena et al.'s sample for
other relatively high proper motion stars that may have been involved
in the ejection event: none were found.  Assuming BN is the only
ejected star and the initial system had negligible velocity with
respect to the ONC, then BN's mass is $m_{\rm BN} = 6.35\pm1.0
(m_{\theta^1C}/50\sm)\sm$. A $\pm0.7$~mas/yr uncertainty in the
initial motion of the system translates into an additional $\pm 2\sm$
uncertainty in $m_{\rm BN}$. BN is now $\sim10$\arcsec\ away
from source ``I'', causing a velocity gain of $\sim4.0
(m_I/40\sm)^{1/2}\kms$. Accounting for this boosts BN's mass
estimate by $\sim10$\%. Treating BN as a massless test particle, its
deflection angle due to ``I'' is $1.56^\circ
(m_{I,*}/20\sm)(b/1000{\rm AU})^{-1}(v_{\rm BN}/36\kms)^{-2}$, where
$b$ is the initial impact parameter and we now normalize to the
expected protostellar mass of ``I'' (Tan 2003).

BN passed close to source ``I'' about $500$~yr ago. Our favored
interpretation is that BN did not form from the gas associated with
``I'' and the hot core (or Kleinmann-Low nebula). This possibility was
first suggested by Zuckerman (1973). However, we cannot rule out ejection
from ``I''.  A close passage of BN with source ``I'' and its
protostellar disk may have led to enhanced angular momentum transport
and accretion via tidal torques (e.g.  Ostriker 1994).
%; Pfalzner 2003).
The protostellar outflow rate is predicted to be proportional to
the accretion rate in a range of theoretical models (e.g. Shu et al.
2000; K\"onigl \& Pudritz 2000).  Allen \& Burton (1993) have argued
that the outflow from ``I'' (i.e.  from the Kleinmann-Low
nebula) shows characteristics consistent with an ``explosive'', i.e.
impulsive, event occurring about one flow crossing time ago ($\sim
10^3\:{\rm yr}$).

%Consider the time-reversed process of the interaction of BN with the
%$\theta^1C$ system: 
For a 3-body interaction, the kinetic energy of BN and $\theta^1C$
%\\
($9.1\times 10^{46} (m_{\rm BN}/7\sm) (v_{\rm BN}/36\kms)^2 \:{\rm
  ergs} + 1.24\times 10^{46}(m_{\theta^1C}/50\sm)(v_{\rm
  \theta^1C}/5\kms)^2\:{\rm ergs}$) should not be significantly
greater than the binding energy of the binary
%\\ 
($1.16\times
10^{47}(m_1/45\sm)(m_2/5\sm)\:{\rm ergs}$, assuming $e=0$).  This
comparison suggests that either the binary is quite eccentric (e.g.
Schertl et al. 2003) or $\theta^1C$'s component masses are larger than
the adopted values.  The period of the $\theta^1C$ binary is about 10
years, so about 400 orbits have occurred since the ejection. Given a
sufficiently accurate determination of BN and
$\theta^1C$'s motion and the $\theta^1C$ binary orbit, details of
the ejection could be constrained.

%Also it suggests that mass of BN is somewhat
%smaller than that inferred from the luminosity of the system
%(\S\ref{S:intro}), as also implied by the momentum of the system.
%Alternatively, we may be dealing with a 4-body interaction: the
%initial state would have been two binaries and there another (unknown)
%runaway star has been produced.

%BN appears (at least in projection) to have passed relatively close to
%the massive protostar, source ``I''. We have argued that the mass
%currently associated with ``I'' is $\sim 40\sm$ ($\sim 20\sm$ in the
%protostar and a comparable amount in the gas envelope, Tan 2003).
%Treating BN as a massless test particle and the protostellar mass of
%``I'' as being concentrated in a point, the deflection angle of the
%orbit of BN with initial impact parameter $b$ is $1.2^\circ
%(M_I/20\sm)(b/1000{\rm AU})^{-1}(v_{\rm BN}/41\kms)^{-2}$. Deviations
%of order this size may be expected in the alignment of the motions of
%$\theta^1C$ and BN, and such deviations, if found, could help to probe
%the mass content of source ``I''. 

\section{The Stellar Wind Bow Shock}\label{S:wind}

A star emitting a wind with mass-loss rate $\dot{m}_w$ and terminal
speed $v_w$ and moving through a medium of density $\rho_a$
%$\rho_a=1.67\times 10^{-24} (1/0.7) n_{\rm H,a}\:{\rm g\:cm^{-3}}$ 
with velocity $v_*$ produces a bow shock at distance $r_{\rm bs}$ in
the direction of motion. Assuming no penetration of ambient
material into the wind bubble and balancing ram pressures, $\rho_w
v_w^2 = \dot{m}_w v_w/(4\pi r_{\rm bs}^2) = \rho_a v_*^2$, we have
\beqa
\label{rho}
n_{\rm H,a} & = & 6.48\times 10^{4} \left(\frac{\dot{m}_w}{10^{-7}\smyr}\right) \left(\frac{v_w}{10^3\kms}\right)\nonumber\\
 & \times & \left(\frac{v_*}{40 \kms}\right)^{-2} \left(\frac{r_{\rm bs}}{300\:{\rm AU}}\right)^{-2}\:{\rm cm^{-3}},
\eeqa
assuming a hydrogen mass fraction of 0.7.

\begin{figure}[h]
\begin{center}
\epsfig{
%prop3
        file=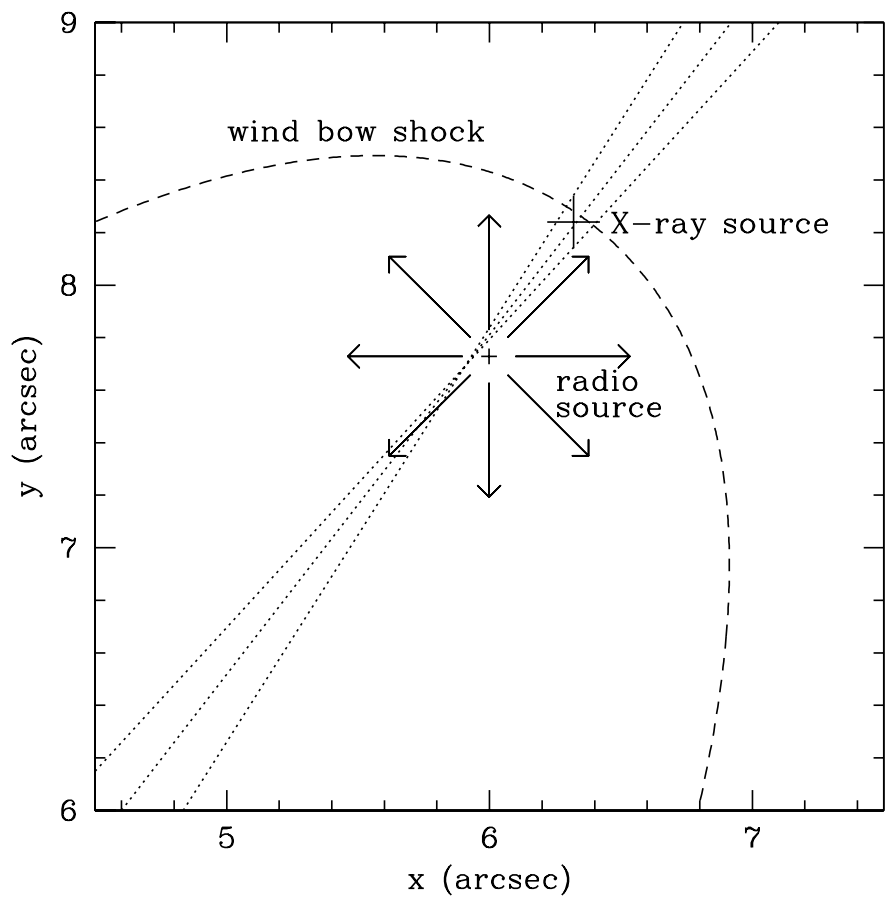,
        angle=0,
        width=\figwidth
}
\end{center}
\caption{ \label{fig:prop5} 
  Offset of X-ray emission (Feigelson et al. 2004, private
  comm.) from BN (Jan 2003, Plambeck, private comm.). The
  coordinates are relative to source ``I''. The size of the crosses
  indicate positional uncertainties. Dotted lines show past and future
  trajectories of BN, as in Fig.~\ref{fig:prop3} (normalized to pass
  through the mean position of the data from 1986.3 to 2004: hence
  slight offset from 2003 position). The dashed line shows the
  projected extent of the bow shock solution of Wilkin (1996) for a
  minimum offset of $r_{\rm bs}=300$~AU (0.67\arcsec).  }
\end{figure}

%The {\it Chandra} X-ray Observatory has reported
X-ray emission from BN
is offset by $\lesssim 1\arcsec$ (Garmire et al. 2000; Feigelson et
al. 2002; Feigelson and the Chandra Orion Ultradeep Project
[COUP] 2004, private comm.). The most recent determination
from COUP is offset from the 2003 radio position by
$0.60\pm0.1\arcsec$ towards P.A. $-32\pm14^\circ$ (Fig.
\ref{fig:prop5}). The density of X-ray sources in this region is $\sim
0.02$ arcsec$^-2$, so the probability of finding an unrelated
source within 1\arcsec of BN is $\sim 0.06$. The alignment of the
observed source with BN's direction of motion suggests that it may be
produced at the wind bow shock.
%% The star to bow shock separation may be estimated from high resolution
%% X-ray observations of BN: for a fast enough wind (see below) we expect
%% X-ray emission from the shocked wind region, which is just inside the
%% shell of shocked interstellar gas.  The {\it Chandra} X-ray
%% Observatory has observed the Orion region on several occasions
%% (Garmire et al. 2000; Feigelson et al.  2002). BN was detected as a
%% hard X-ray source, consistent with being heavily absorbed. Garmire et
%% al. (2000) reported an offset of about $1.1\pm0.5\arcsec$ of the X-ray
%% source from the radio source towards the northwest. With approximately
%% twice the amount of data, Feigelson et al. (2002) reported a smaller
%% offset, equivalent to $0.34\arcsec$ towards position angle $-39^\circ$
%% from the 2003 radio position. The Chandra Orion Ultradeep Project
%% (COUP, Feigelson et al., in prep.) has about 13~days of data.  The
%% position of the X-ray source from this data is shown in Figure
%% \ref{fig:prop5} and is accurate to about $\pm 0.1\arcsec$. We find the
%% offset from the 2003 radio position to be $0.60\pm0.1\arcsec$ towards
%% position
%% angle $-32\pm14^\circ$.%${^{-16^\circ}_{-44^\circ}}$. 
%% %Note that while the position of the source is known
%% %to about $\pm 0.1\arcsec$, the size
If the X-rays come from the inner edge of the thin shell of
shocked wind and ambient gas, with each solid angle of wind material
from a region about 1\arcsec across contributing equally, then we can
use the solution of Wilkin (1996) for the shell geometry to
estimate $r_{\rm bs}$.
%In practice we only include
%contributions from wind material from the forward-moving hemisphere.
The offset of the center of the X-ray emission is 90\% of
$r_{\rm bs}$. For the fiducial geometry, projection effects increase
it by $\sim 5\%$, so we estimate $r_{\rm bs}=300\pm50$~AU.  We note
that in this interpretation the bow shock is much further from the
star than the extent of the ionized region, which is tens of AU. The
stellar ionizing flux is confined within the wind even before it
reaches the bow shock (c.f. Van Buren et al. 1990).
%% The position angle of the offset X-ray emission
%% is in approximately the same direction as the proper motion of the
%% source (Figure \ref{fig:prop5}).  It is also almost diametrically
%% opposite to the direction to $\theta^1C$.  While these facts support
%% the interpretation that the X-ray source traces the stellar wind
%% bow shock, we must keep in mind the possibility that the source is
%% simply unrelated to BN. 

The X-ray spectrum constrains the initial wind velocity,
since the temperature of the postshock ($\gamma=5/3$) gas is $kT =
(3/16)\mu v_s^2 = 1.96 (\mu/m_{\rm H}) (v_s/10^3\kms)^2\:{\rm keV}$,
and the shock speed $v_s\simeq v_w$. Based on 45 counts,
Feigelson et al. (2002) reported that the emission is hard ($kT>10$~keV)
and heavily absorbed ($N_{\rm H}\sim4\times 10^{22}\:{\rm cm^{-2}}$).
The observed (0.5-8~keV) luminosity is $2.5\times 10^{29}\:{\rm
  ergs\:s^{-1}}$, becoming $\sim 4\times 10^{29}\:{\rm ergs\:s^{-1}}$ once
corrected for absorption. With so few counts the
uncertainties in these parameters are large: $\sim 50\%$
in $kT$ and a factor of $\sim$3 in $N_{\rm H}$. 
%More
%accurate estimates will be provided by the forthcoming analysis of the
%COUP data (Feigelson et al., in prep.). 
Nevertheless these
results suggest the presence of a fast wind. The escape speed
from main sequence stars with $m_*=8.4-12\sm$ (described in
\S\ref{S:intro}) are $v_{\rm esc,*}=970-1030\kms$.
Consistency with the observed X-ray spectrum may be possible for
a model with a somewhat lower temperature (i.e. $\sim$few keV),
an increased amount of absorption, and a much higher luminosity (see below).
%, more consistent with the
%interpretation of Bunn et al. (1995) that $A_V>18$~mag.

The wind bow shock model predicts a relatively constant X-ray
source, unless there are large variations in the power of the stellar
wind or the density of the ambient medium. The limited variability
reported by Feigelson et al. (2002) is consistent with a constant
source. The model also predicts that the source may show extended
emission on scales $\sim1$\arcsec\, but perhaps of low surface
brightness.

The mass-loss rates from young B stars are uncertain. Even for
stars on the main sequence the theoretical mass-loss
rate due to a line-driven wind is uncertain because the star may be
close to the ``bi-stability jump'' at $T_*\simeq 21000\:{\rm K}$,
where there is a change in the ionization state of the lower wind
layers near the sonic point (Vink et al. 1999). Using the results of
Vink et al. (2001), the $m_*=8.4\sm$ case can have $\dot{m}_w\sim 2 -
70 \times 10^{-11} \smyr$, with $v_w\simeq 2.7 - 1.3 v_{\rm esc,*}$.
For the $m_*=12\sm$ case $\dot{m}_w=1.15\times 10^{-9}\smyr$ and
$v_w\simeq 1.3 v_{\rm esc,*}$. 

It takes a time
%\\ 
$t_{\rm KH}\equiv a_g \beta G m_*^2/(2 r_* L_*) =
1.57\times 10^4 a_g \beta (m_*/10\sm)^2 (r_*/10R_\odot)^{-1}
(L_*/10^4L_\odot)^{-1}\:{\rm yr}$ to settle to the main sequence,
where $\beta$ is the mean ratio of gas to total pressure
($\simeq 1$ for $m_*\sim10\sm$) and $a_g\equiv 3/(5-n)$ for polytropes
with $n<5$.  Approximating $n=3$ then $t_{\rm KH}\simeq 20 - 8\times
10^4\:{\rm yr}$ for the $m_*=8.4 - 12\sm$ cases, respectively. Thus it
is possible
%, particularly for the lower mass estimates of BN, 
that the star is still in a pre-main sequence phase. If so, then
empirically we expect higher mass-loss rates, $\sim 10^{-8} -
10^{-6}\smyr$ (Nisini et al. 1995) and somewhat lower wind velocities
(by factors of a few), compared to main sequence wind models.  Indeed
models of slower, denser winds have been proposed to explain radio
continuum emission and near-infrared hydrogen recombination lines
observed from BN. Scoville et al. (1983) estimated $\dot{m}_w\simeq
4\times 10^{-7}\smyr$ and $v_w>100\kms$. H\"oflich \& Wehrse (1987)
presented a non-LTE model with $\dot{m}_w=3\times 10^{-7}\smyr$ and
$v=20\kms$. Bunn et al. (1995) observed Br~$\gamma$ line wings
extending to $\sim220\kms$.
%We note that the lack of emission on the blueward side may
%be due to absorption from blueshifted gas in the outflow from source
%``I''. 
%, which would require BN to be somewhat behind source ``I''.
%Scoville et al. (1983) also noted blueshifted absorption.
Note that the wind luminosity is $L_w=3.15\times
10^{34}(\dot{m}_w/10^{-7}\smyr)(v_w/1000\kms)^2\:{\rm ergs\:s^{-1}}$
($=8.2L_\odot$), which is small compared to the bolometric luminosity,
but large compared to the X-ray luminosity.

%% Estimates of the mass-loss
%% rate from a star with $L=2500\:L_\odot$, $m_*=8.4\sm$,
%% $r_*=3.4R_\odot$ and $T=22000\:{\rm K}$ range from $\sim 2\times
%% 10^{-11}\smyr$ to $\sim 7\times 10^{-10}\smyr$ (Vink et al. 2001). For
%% the lower value, the wind speed is expected to be about 2.7 times
%% $v_{\rm esc,*}=970\kms$, while for the upper value it is 1.3 times
%% this. For the higher mass case: $L=10^4\:L_\odot$, $m_*=12\sm$,
%% $r_*=4.3R_\odot$ and $T=28000\:{\rm K}$ we have $v_w=2.7v_{\rm esc}$,
%% $v_{\rm esc,*}=1030\kms$ and $\dot{m}_w=1.15\times 10^{-9}\smyr$.
%% However, even the larger of these values may be a significant
%% underestimate because of the youth of the star (ref??? - how good are
%% these models for $\theta^1C$? -schultz???).

Given the uncertainties in $\dot{m}_w$ and $v_w$, we cannot use the
bow shock model to estimate unambiguously the ambient density.
However, the product of these three quantities is constrained
(eq.~\ref{rho}). Estimates of the mean density (e.g. \S\ref{S:motion})
are about an order of magnitude higher that the fiducial value shown
in eq.~\ref{rho}, suggesting (in the context of our bow shock model)
that $\dot{m}_w$ and/or $v_w$ are relatively large compared to the
adopted fiducial values.

We note that some of the H recombination line emission that
motivated ``slow'' wind models may be generated by shocked ambient
interstellar material.  From mass continuity of gas entering the
shocked shell and flowing around the wind bubble, the shell's column density
is approximately $n_{\rm H,a} r_{\rm bs} = 4.5\times
10^{20} (n_{\rm H,a}/10^5\:{\rm cm^{-3}})(r_{\rm bs}/300\:{\rm
  AU})\:{\rm cm^{-2}}$. As an aside, this is small compared to the
total inferred absorbing column, which must be dominated by
intervening ambient material. For an isothermal shock with
$T_a=100\:{\rm K}$ and $\mu_a=2.35m_{\rm H}$ so that $c_a=0.59\kms$,
then the Mach number is $\sim70$ and the compression ratio $\sim 5000$
for $v_*=40\kms$. Thus an upper limit to the shocked density is $\sim
10^{10}\:{\rm cm^{-3}}$, assuming $n_{\rm H,a}\sim
10^6\:{\rm cm^{-3}}$. The actual densities may be much smaller if
magnetic fields contribute significantly to the total pressure.  

From CO emission features, Scoville et al. (1983) reported the
presence of molecular gas with $n_{\rm H_2}\sim 10^7 -10^{12}\:{\rm
  cm^{-3}}$ and $T\sim 600-3500\:{\rm K}$. Some of this emission may
be due to the swept-up shell, but the highest densities are difficult
to explain with the bow shock model. A more likely alternative is that
the densest gas, $\lesssim 1$~AU from the star, is the remnant of an
accretion disk. This is the favored interpretation to explain CO
emission from a number of luminous young stellar objects (e.g. Ishii et al.
2001). The disk is probably not being fed
significantly by swept-up material: the Bondi-Hoyle accretion rate,
ignoring stellar feedback, is still relatively small,
$\dot{m}_{\rm BH} = 1.3\times 10^{-9} (n_{\rm H,a}/10^5\:{\rm
  cm^{-3}}) (m_*/10\sm)^2 (v_*/40\kms)^{-3}\smyr$ (e.g. compared to
the expected mass loss rate). If, however, the ambient medium is
clumped on very small scales then it would be easier to occasionally
bring dense interstellar molecular gas close to BN at higher rates.

\acknowledgements JCT is supported by a Spitzer-Cotsen fellowship from
Princeton Univ. and NASA grant NAG5-10811. We thank R. Plambeck
and E. Feigelson for sharing unpublished data on BN. We also thank B. Draine, L. Eyer, J.
Goodman, L. Hillenbrand, P. Hut, C. McKee, C. R. O'Dell, W. van Altena, W.-J.
de Wit and the referee for discussions.

%--------------------------------------------------------------------------
\end{document}